\documentclass[aps,twocolumn,showpacs,prl]{revtex4}
\usepackage[dvips]{graphicx}
\usepackage{latexsym}
\usepackage{amsmath}

\newcommand{\BE}{\begin{equation}}
\newcommand{\EE}{\end{equation}}
\newcommand{\BA}{\begin{eqnarray}}
\newcommand{\EA}{\end{eqnarray}}

\begin{document}

\title{Self-consistent variational approach to the minimal left-right symmetric 
model of electroweak interactions}

\author{Fabio Siringo}
\author{Luca  Marotta}
\affiliation{Dipartimento di Fisica e Astronomia, 
Universit\`a di Catania,\\
INFN Sezione di Catania and CNISM Sezione di Catania,\\
Via S.Sofia 64, I-95123 Catania, Italy}

\date{\today}
\begin{abstract}
The problem of mass generation is addressed by a Gaussian variational method for the minimal
left-right symmetric model of electroweak interactions. Without any scalar bidoublet, the Gaussian
effective potential is shown to have a minimum for a broken symmetry  vacuum with a finite
expectation value for both the scalar Higgs doublets. The symmetry is broken by the fermionic
coupling that destabilizes the symmetric vacuum, yielding a self consistent fermionic mass.
In this framework a light Higgs is only compatible with the existence of
a  new high energy mass scale below 2 TeV.
\end{abstract}

\maketitle

\section{introduction}

It has been recently suggested\cite{siringo,brahmachari,siringo2}
that mass generation and the breaking of left-right symmetry can be
described by a minimal left-right symmetric model with only two scalar doublets and no
bidoublets. Several left-right symmetric extensions of the standard model have been developed by
many authors\cite{pati,mohapatra,senjanovic2,senjanovic} 
and are based on the gauge group  $SU(2)_L\times SU(2)_R\times U(1)_{B-L}$.
They have the remarkable merit of predicting the same low energy phenomenology of the standard model\cite{theorem},
thus explaining the lack of symmetry of  the electroweak interactions. However, most of those models also contain
a scalar Higgs bidoublet and require the existence of ten physical Higgs particles at least. A minimal model with
only two scalar doublets  $\Phi_L$, $\Phi_R$ transforming as $(2,1,1)$ and $(1,2,1)$ respectively, would
require the existence of just two physical Higgs particles and would be an appealing alternative provided that
a stable vacuum with a non zero expectation value $v_L=\langle\Phi_L\rangle$ can be found.
It has been pointed out\cite{siringo2,senjanovic2} that without any bidoublet the effective potential 
has a minimum for $\langle\Phi_R\rangle=v_R\not=0$ and $\langle\Phi_L\rangle=v_L=0$ and the model would
be useless as $v_L=v_{Fermi}$ gives the scale of all the known particle masses. In fact the insertion of a
scalar bidoublet was the simplest way of shifting the minimum of the effective potential towards a small but finite 
$v_L$ value. In spite of that, the minimal model has gained some success and its prediction of  a new
intermediate physical mass scale $v_R$ has been already explored\cite{almeida}showing that
the  new physics could appear at the TeV scale in the new electron-positron colliders.

In this paper we give a self-consistent solution to the open problem\cite{siringo2} 
of a physical vacuum for the minimal model: we show that the inclusion of quantum fluctuations yields a
stable vacuum with $v_L\not= 0$ even without any bidoublet, thus motivating and enforcing previous
work on the model\cite{brahmachari,almeida}.
The problem  is addressed by a variational method, as we show that
in the minimal model the minimum of the Gaussian  Effective 
Potential\cite{schiff,rosen,barnes,kuti,chang,weinstein,huang,bardeen,peskin,stevenson}
is shifted towards a small finite $v_L$ value by the
coupling to fermions. The variational character of the calculation is important as it ensures that the 
$v_L=0$ vacuum is not stable even without any scalar bidoublet. 
Thus the coupling to fermions is no longer a mere way to generate
the Dirac masses, but it becomes the source of symmetry breaking for the field $\Phi_L$ yielding a
self consistent mass for the heavy top quark.

The paper is organized as follows: in section II an effective low-energy Lagrangian is recovered
from the full left-right symmetric model; the vacuum stability of the effective model is studied in
section III by a variational method for the effective potential; in section IV the effective potential
is discussed together with its predictions for the mass spectra.

\section{effective low-energy model}

The problem of a viable physical vacuum for the minimal model has been recently addressed by invoking
the existence of a larger approximate global symmetry in addition to the discrete parity symmetry\cite{siringo}.
The physical Higgs particle would emerge as the pseudo-Goldstone boson associated with the breaking of
the global symmetry. In that scenario the fermionic one-loop contributions to the effective potential become
relevant for determining the minimum of the effective potential and the mass scale $v_L$. It emerges that the
zero-point fermionic energy destabilizes the $v_L=0$ vacuum in competition with the bosonic terms.
The approximate global symmetry has the further merit of being accidentally respected by any quadratically
divergent contribution to the Higgs potential\cite{chacko}, thus we expect that the global symmetry should
survive in the low energy effective Lagrangian of the Higgs sector up to small symmetry breaking terms.
Then we take the Higgs potential to be
\BA
{\cal L_H}&=&\frac{1}{2}M_B^{2}\left(\Phi_L^\dagger\Phi_L+\Phi_R^\dagger\Phi_R\right)+
\nonumber\\
&+&\frac{1}{4!}\lambda \left(\Phi_L^\dagger\Phi_L+\Phi_R^\dagger\Phi_R\right)^2+
\frac{1}{2} \eta \>\Phi_L^\dagger\Phi_L\Phi_R^\dagger\Phi_R
\label{LH}
\EA
where $\eta$ is a small parameter which breaks the global $SU(4)$ symmetry. 
At variance with soft symmetry breaking models\cite{babu} where the left-right symmetry is
broken by insertion of different masses for left and right fields, our Higgs potential is assumed to be fully
symmetric. Thus the present model provides a truly spontaneous symmetry breaking mechanism.

In unitarity gauge the doublets may
be taken as $\Phi_{L,R}=(0, \phi_{L,R})$ where the scalar real components $\phi_{L,R}$ describe
two physical neutral bosons. For $\eta>0$ and $M_B^2<0$ the potential has a minimum for $\phi_L=0$ and
$\phi_R^2=v_R^2=-6M_B^2/\lambda$. We may set $\phi_R=v_R+h_R$ and write the new potential for
the field $h_R$ as ${\cal L_H}={\cal L}_L+{\cal L}_R+{\cal L}_{int}$
\BE
{\cal L}_L=\frac {1} {2} (\eta v_R^2)\phi_L^2+\frac{1}{4!}\lambda \phi_L^4
\label{LL}
\EE
\BE
{\cal L}_R=\frac {1} {3!} (\lambda v_R^2)h_R^2+\frac{1}{3!}(\lambda v_R)h_R^3
+\frac{1}{4!}\lambda h_R^4
\label{LR}
\EE
\BE
{\cal L}_{int}=\frac {1} {2} \left(\eta+\frac{\lambda}{3!}\right) \phi_L^2 h_R^2
+\left(\eta+\frac{\lambda}{3!}\right) v_R\phi_L^2h_R
\EE

Assuming that $\lambda\gg\eta$, the heavy Higgs field $h_R$ is decoupled and we can regard
Eq.(\ref{LL}) as the effective low-energy Lagrangian for the field $\phi_L$. A natural cut-off
$\Lambda\approx v_R$ comes from the requirement that the physical vacuum is stable around
the broken symmetry point $\phi_L=0$, $\phi_R=v_R$.
The full effective Lagrangian for the field $\phi_L$ is easily recovered by adding the kinetic terms and
the coupling to fermions and gauge bosons. 
Provided that the vacuum expectation value $\langle \phi_L\rangle=v_L$ does
not vanish, the model gives rise to the standard phenomenology with the gauge bosons acquiring masses
at two different scales: a small mass scale $v_L$ for the standard model  light bosons $W^{\pm}_L$,
$Z$, and a larger mass scale $v_R$ for the heavy gauge bosons $W^{\pm}_R$, $Z^\prime$.
The details of the symmetry breaking are dicussed in Ref.\cite{siringo} where the standard model Lagrangian
is recovered up to ${\cal O}(v_L^2/v_R^2)$ corrections. Moreover, without any Higgs bidoublet, 
the charged bosons  $W^{\pm}_L$ and $W^{\pm}_R$ are shown to be exactly decoupled. 
The heavy bosons are decoupled any way in the low energy domain where they hardly play any role if
$v_R\gg v_L$.

Here our main concern is the shift of the vacuum expectation value of $\phi_L$ from the $\phi_L=0$
minimum of the potential ${\cal L}_L$ in Eq.(\ref{LL}). Thus we neglect the weak coupling to the gauge
bosons and insert the coupling to Dirac fermion fields $\psi_i$\cite{siringo,brahmachari,weinberg}
\BE
{\cal L}_f=\sum_i q_i \bar\psi_i\psi_i\phi_L\phi_R
\label{Lf}
\EE
where the index $i$ runs over the different kinds of quarks and leptons. Since 
all the fermions but the top quark $\psi_t$ may be neglected for their small coupling constants,
in the Euclidean formalism the full effective Lagrangian reads
\BE
{\cal L}_{eff}=\frac{1}{2} (\partial_\mu \phi_L)^2+\bar\psi_t(\rlap/\partial+\alpha\phi_L)\psi_t+
\frac {1} {2}M^2 \phi_L^2+\frac{1}{4!}\lambda \phi_L^4
\label{efflag}
\EE
where
$M^2=\eta v_R^2>0$ and $\alpha=q_t v_R$.

A simple dimensional argument would suggest that the coupling constants $q_i$ scale like the inverse of some
large energy $\Gamma$.  Moreover, according to Eq.(\ref{efflag}), the term $\alpha v_L$ is the
mass of the top quark $m_t$ which is not too much smaller than $v_L$. Thus the adimensional
coupling $\alpha$ is of order unity and
$\Gamma$ turns out to be only slightly above $v_R$. One may wonder in that case whether the effective operator
in Eq.(\ref{Lf}) makes sense for the top quark. The problem has been addressed in Ref.\cite{brahmachari}
where it is shown that up to a  renormalization of couplings no serious change occurs in mass spectra.
Of course, in order to address this point further, some extra hypothesis would be required on the origin of the
dimension-five operator\cite{siringo,brahmachari,weinberg} in Eq.(\ref{Lf}).

At tree level the effective Lagrangian Eq.(\ref{efflag}) predicts a vanishing vacuum expectation value $v_L$ for
the field $\phi_L$. However, in the next section we show that quantum fluctuations 
make the $v_L=0$ vacuum unstable
towards a phenomenologically acceptable finite value.

\section{The Gaussian Effective Potential}

It has been recently shown\cite{marotta,camarda} that in three dimensions the scalar theory can be studied by the 
Gaussian Effective Potential 
(GEP)\cite{schiff,rosen,barnes,kuti,chang,weinstein,huang,bardeen,peskin,stevenson0,stevenson} 
yielding a very good interpolation of the experimental correlation  lengths of superconductors. As those lengths
are the inverse of the masses, we believe that the GEP method could give a reliable estimate of masses
in four dimensions as well. The variational nature of this approximation makes it a more powerful tool 
than any perturbative approach, such as the one-loop effective potential (1LEP); in fact, it has been 
shown (see, for instance \cite{kuti, chang, stevenson0}) that the GEP contains features which can not be 
obtained by calculations based on a finite order of loops, and it results well-defined even when the 
1LEP becomes a complex (and so unphysical) quantity. Moreover the variational character of the GEP
makes it reliable even in the strong coupling limit where the 1LEP makes no sense. In fact
the effective Lagrangian Eq.(\ref{efflag}) contains a Higgs self-interaction parameter $\lambda$
which is known to be of order unity or larger (\cite{siringo3}), and any perturbative prediction on the vacuum 
stability would not be conclusive.

The GEP for the Lagrangian (\ref{efflag}) has been studied by several authors\cite{stevenson2,stancu2} and
it is well known to be unbounded: the fermions encourage spontaneous symmetry breaking, and
in four dimensions they destabilize the scalar theory in the limit of an infinite cut-off $\Lambda\to \infty$.
However the effective theory has a natural cut-off $\Lambda\approx v_R$ and the inclusion of the fermionic
couplings just gives rise to a spontaneous symmetry breaking for the field $\phi_L$ which acquires a small finite
expectation value $v_L\ll v_R$. 

As usual for the GEP method\cite{stevenson,stancu} we take 
$\phi_L=\varphi +h_L$ where $\varphi$ is a constant shift and $h_L$ is the Higgs field. The Gaussian
Lagrangian is taken as
\BE
{\cal L}_{GEP}=\frac{1}{2} (\partial_\mu h_L)^2+
\frac {1} {2}\Omega^2 h_L^2+
\bar\psi_t(\rlap/\partial+m)\psi_t
\label{geplag}
\EE
where $\Omega$ and $m$ are variational parameters for the masses. The GEP follows\cite{stevenson2}
\BA
V(\varphi)&=&\left[I_1(\Omega)-4 I_1(m)\right] +\frac{1}{2} M^2 \varphi^2+\frac {1}{4!}\lambda\varphi^4+
\nonumber\\
&+&\left[4m(m-\alpha\varphi) I_0(m)\right]+\nonumber\\
&+&\frac{1}{2} I_0(\Omega)\left[M^2-\Omega^2+
\frac{1}{2}\lambda\varphi^2+\frac{1}{4}\lambda I_0(\Omega)\right]
\label{gep}
\EA
where the integrals $I_n (x)$ are defined acording to
\BE
I_1(x)=\frac{1}{2}\int \frac{d^4p}{(2\pi)^4}\log (p^2+x^2)
\label{I1}
\EE
\BE
I_0(x)=2\frac{dI_1(x)}{dx^2}=\int \frac{d^4p}{(2\pi)^4}\frac{1}{p^2+x^2}
\label{I2}
\EE
These diverging integrals are supposed to be regularized by the cut-off $p<\Lambda$, and their
exact evaluation is trivial.

The minimum of $V(\varphi)$  is found by solving the system of coupled equations 
$\partial V/\partial \Omega=\partial V/\partial m=\partial V/\partial \varphi=0$, which reads
\BA
\Omega^2&=&M^2+\frac{1}{2}\lambda\varphi^2+\frac{1}{2}\lambda I_0(\Omega)
\label{gap1}\\
m&=&\alpha\varphi
\label{gap2}\\
4\alpha m I_o(m)&=&\varphi\left[M^2+\frac{1}{3!}\lambda\varphi^2+\frac{1}{2}\lambda I_0(\Omega)\right]
\label{min}
\EA
According to Eq.(\ref{gap2}) the fermionic contribution to the effective potential is just the first order 
perturbative term evaluated at the self consistent mass $m=\alpha\varphi$.

\begin{figure}[ht]
\includegraphics[height=9cm, width=9cm]{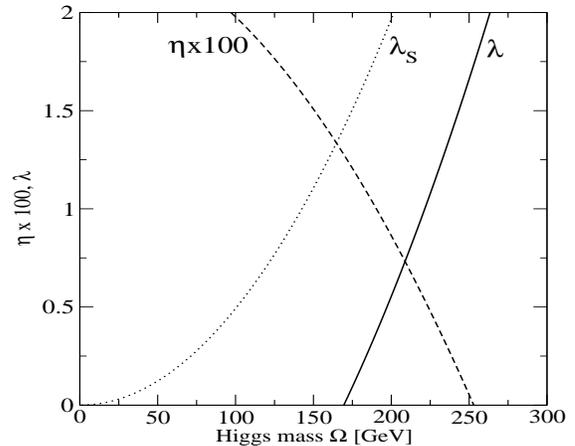}
\caption{\label{Fig1}
The free parameters $\lambda$ (solid line) and  $\eta$ (dashed line) 
evaluated by Eqs.(\ref{min2}),(\ref{eta}) as functions
of the Higgs mass $\Omega$ at the scale $v_R=1500$ GeV ($v_R/v_L\approx 6$).
Since $\eta\ll\lambda$, the parameter $\eta$ has been rescaled by a factor 100.
For comparison the standard\cite{stevenson} GEP  $\lambda_S$ is also reported (dotted line).
The constraint $\lambda,\eta>0$ is only satisfied for $170<\Omega<253$ GeV for this choice of $v_R$.}
\end{figure}
We can show that for a broad range of the parameters the coupled equations have a broken symmetry solution
for $\varphi=v_L\not=0$, thus yielding a finite mass for the fermion $m=\alpha v_L$. That can be easily seen  
by fixing  $v_L$,$v_R$, $\Omega$ and $m$ at reasonable phenomenological values and by
solving the coupled equations (\ref{gap1}),(\ref{min}) in order to get the free parameters $\lambda$, $\eta$.
Since we want  to recover the correct weak interaction phenomenology, 
the expectation value of $\phi_L$ must be set equal to the Fermi value $v_L=v_{Fermi}=247$ GeV. 
Moreover the Top quark mass is known to be $m_t=181$ GeV and then
the coupling must be $\alpha=m_t/v_L=0.733$.
By use of  Eq.(\ref{gap1}) we can write Eq.(\ref{min}) as
\BE
\lambda=\frac{3}{v_{Fermi}^2}\left[\Omega^2-4\alpha^2 I_0(m_t)\right]
\label{min2}
\EE
and this already provides a lower bound on the Higgs mass $\Omega$ as for $x\ll \Lambda$, 
$I_0(x)\approx \Lambda^2/(16\pi^2)$ and the existence of a solution $\lambda>0$
requires that $\Omega>(\alpha\Lambda)/(2\pi)\approx 0.12\Lambda$. Thus the new high energy
mass scale $v_R\approx\Lambda$ cannot be larger than ten times the Higgs mass $\Omega$.
A comparison with the standard\cite{stevenson} 
GEP relation $\lambda_S=3\Omega^2/v^2_{Fermi}$ shows
that the parameter $\lambda$ is smaller in this minimal model, thus allowing for perturbative
tretments up to a bit larger Higgs mass.
However, as it can be seen from Fig. 1, the coupling costant $\lambda$ comes out to be of order of unity even 
for Higgs mass values $\Omega$  quite close to the experimental lower bound; this circumstance makes clear 
our choice of using the GEP variational method: the problem we are dealing with is a non perturbative one
and an analysis based on perturbation theory is not reliable.
Eq.(\ref{gap1}) reads
\BE
\eta=\frac{1}{v_R^2} \left[ \Omega^2-\frac{1}{2}\lambda\left(v^2_{Fermi}+I_0(\Omega)\right)\right]
\label{eta}
\EE
The existence of a broken symmetry minimum in the potential ${\cal L_H}$ Eq.(\ref{LH})
requires that $\eta>0$, and since $\lambda=3\Omega^2/v^2_{Fermi}+const.$, according to Eq.(\ref{min2}) ,
we find an upper bound for the Higgs mass. At any fixed high energy scale $v_R\approx\Lambda$, the
dependence on $\Omega$ of the parameters $\lambda$, $\eta$ is obtained by Eqs.(\ref{min2}),(\ref{eta}).
We notice that, apart from a very narrow range at the lower bound of $\Omega$, the parameter $\eta$ is always
small and $\eta\ll \lambda$ as it was required in order to decouple the heavier Higgs field $h_R$.

\section{discussion}

The existence of a broken symmetry vacuum with $v_L\not=0$  is a variational result: as the exact vacuum must
have a smaller energy, we can conclude that the $v_L=0$ vacuum cannot be stable. Thus the choice of a 
variational method gives us more confidence on its prediction of a broken symmetry vacuum. 
Moreover the GEP gives rise to some bounds for the Higgs boson mass according to Eqs.(\ref{min2}),(\ref{eta}).
The general behaviour is shown in Fig.1 for $v_R=1500$ GeV ($v_R/v_L\approx 6$) where
the constraint $\lambda,\eta>0$ is only satisfied for $170<\Omega<253$ GeV. 

\begin{figure}[ht]
\includegraphics[height=10cm, width=10cm]{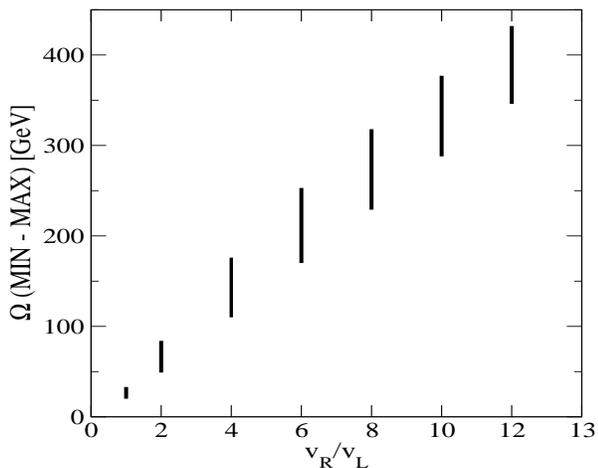}
\caption{\label{Fig2}
The allowed range for the Higgs mass $\Omega$ (vertical bars) 
are reported for several choices of the ratio  $v_R/v_L$ between high energy
and low energy mass scales ($v_L$ is fixed at the Fermi value $v_{Fermi}=247$ GeV).}
\end{figure}

\begin{figure}[ht]
\includegraphics[height=12cm, width=10cm,angle=-90]{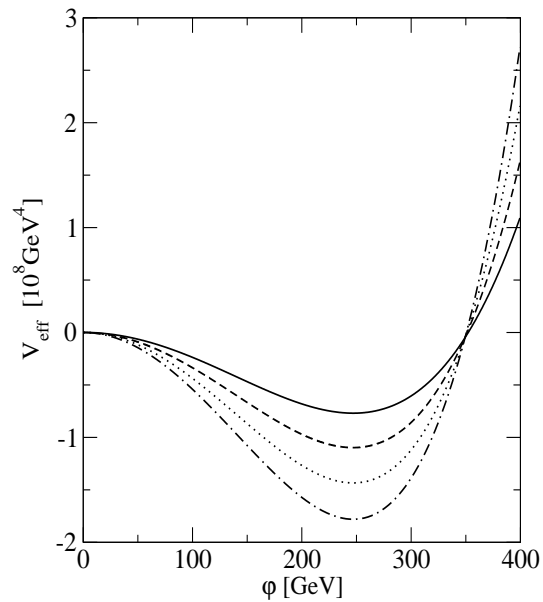}
\caption{\label{Fig3}
The Gaussian effective potential according to Eq.(\ref{gep}) for several choices of the Higgs mass and
the high energy scale $v_R$. From the top: 
$\Omega=140$ GeV, $v_R=1000$ GeV (solid line);
$\Omega=200$ GeV, $v_R=1500$ GeV (dashed line); 
$\Omega=260$ GeV, $v_R=2000$ GeV (dotted line);
$\Omega=320$ GeV, $v_R=2500$ GeV (dot-dash).
For each value of the shift $\varphi$ the GEP
is evaluated by inserting in Eq.(\ref{gep}) the self-consistent masses $\Omega$, $m$ which solve the
gap equations Eqs.(\ref{gap1}),(\ref{gap2}). 
The minimum point $\varphi=v_L$ does not change as it is fixed at the 
phenomenological value $v_{Fermi}=247$ GeV}.
\end{figure}

The allowed range for the Higgs mass $\Omega$ is reported in Fig.2 for several choices of the high energy
scale $v_R$.
Assuming a rather light Higgs mass $\Omega\approx 200$ GeV, then the model is only consistent with an
high energy scale $v_R\leq 1.7$ TeV. If for instance we take $v_R=1.5$ TeV, then 
the coupled linear equations (\ref{min2}), (\ref{eta}) 
have the solution $\lambda=0.554$ and $\eta=(M^2/v_R^2)=8.6\cdot 10^{-3}$. 
In this scenario the search for new physics at the TeV scale
could reveal the existence of  the heavy Higgs field $h_R$ and the new heavy gauge bosons $W^{\pm}_R$
and $Z^\prime$, whose masses\cite{siringo} would be larger by a factor $v_R/v_L\approx 6$ 
compared to their light partners.
However, according to Fig.2, a larger Higgs mass would be compatible with a larger high energy scale, and a
quite heavy Higgs at the TeV scale would require a large $v_R\approx 10$ TeV which in turn would push
any chance to reveal the new physics towards higher energies.
The effective potential Eq.(\ref{gep}) is reported in Fig.3 for average Higgs masses at different high energy
scales $v_R$. 
For each value of the shift $\varphi$ the GEP
is evaluated by inserting in Eq.(\ref{gep}) the self-consistent masses $\Omega$, $m$ which solve the
gap equations Eqs.(\ref{gap1}),(\ref{gap2}). 
The minimum point $\varphi=v_L$ does not change as it is fixed at the 
phenomenological value $v_{Fermi}=247$ GeV by solution of the equations Eqs.(\ref{min2}), (\ref{eta})
for the parameters $\lambda$, $\eta$. Of course the curvature does change as the Higgs mass $\Omega$ does.

This simple variational study shows that the minimal left-right symmetric model, with only two Higgs doublets and
no bidoublets could give a satisfactory description of the known phenomenology
provided that quantum fluctuations are included. 
The model predicts the existence
of a high energy mass scale $v_R$ which is not larger than ten times the Higgs mass. At the new energy scale
the three right handed partners of the weak bosons should come out in experiments together with the second
heavier higgs boson $h_R$. The model does not require the existence of other particles, and its  predictions 
on masses could be tested in the new electron-positron colliders. 

In summary we have shown that inclusion of the fermionic coupling is enough for destabilizing the $v_L=0$
vacuum of the minimal left-right symmetric model, thus yielding a small finite low energy scale $v_L\ll v_R$. 

The genuine variational nature of the GEP method ensures that the instability of the $v_L=0$ vacuum is not 
a mere artifact introduced by the approximation, but a property of the model; furthermore the method
allows for reliable predictions even in the strong coupling (heavy Higgs) regime, where any 
perturbative treatment (e.g. 1LEP) would not be valid.
Thus even without any bidoublet, the minimal model is shown to be a viable framework for explaining
the breaking of left-right symmetry in electroweak interactions.


\begin{thebibliography} {99}
\bibitem{theorem} That left-right symmetric theories predict the standard model phenomenology has been shown 
   on very general grounds by Senjanovic\cite{senjanovic} who employed the method of
   Georgi and Weinberg\cite{georgi2}.
\end{thebibliography}
\end{document}